# A simple method to predict the magnetic properties of benzenoid polycyclic hydrocarbons

Yuyi Yan[1], Fengru Zheng[1], Boyu Qie[2,3], Jiayi Lu[1], Hao Jiang[1], Zhiwen Zhu[1], Qiang Sun*[,1]

[1]Materials Genome Institute, Shanghai University, 200444 Shanghai, China

[2]Department of Chemistry, University of California, Berkeley, CA 94720, USA

[3]Kavli Energy NanoScience Institute at the University of California Berkeley and the Lawrence Berkeley National Laboratory, Berkeley, California 94720, USA

**ABSTRACT:** Open-shell benzenoid polycyclic hydrocarbons (BPHs) are promising materials for future quantum applications. However, the search and realization of open-shell BPHs with desired properties is a challenging task due to the gigantic chemical space of BPHs, requiring new strategies for both theoretical understanding and experimental advancement. In this work, by building a structure database of BPHs through graphical enumeration, performing data-driven analysis, and combining tight-binding and mean-field Hubbard calculations, we discovered that the number of the internal vertices of the BPH graphs is closely correlated to their open-shell characters. We further established a series of simple rules, the triangle counting rule (TCR), to predict the magnetic ground state of BPHs. These findings not only provide a database of open-shell BPHs, but also extend the well-known Lieb's theorem and Ovchinnikov's rule and provide a straightforward method for designing open-shell carbon nanostructures. These insights may aid in the exploration of emerging quantum phases and the development of magnetic carbon materials for technology applications.

## INTRODUCTION

Benzenoid polycyclic hydrocarbons (BPHs), also referred to as polycyclic $\pi$-conjugated hydrocarbons or nanographenes, are fragments of graphene with mono-hydrogen passivated edges. BPHs possess unique open-shell properties due to their flexible magnetism endowed by the $p$ or $\pi$ electrons, and are thus considered promising candidates for next-generation electronics and spintronics.[1] The magnetic configurations of open-shell BPHs, which include the spin quantum number $S$ and magnetic exchange interactions, can be predicted by simple approximations and are highly tunable through variations in molecular size, shape and edge structure.[2-19] The exploration of the spin degrees of freedom of unpaired electrons in open-shell BPHs could date back over a century ago[20-21] and continues to be a very active area of research. Despite previous efforts, the synthesis of open-shell BPHs remains challenging. Excitingly, recent advances in on-surface synthesis methods under ultrahigh-vacuum conditions have enabled the synthesis of highly reactive open-shell BPHs including triangulene,[2-5] zethrene,[6-7] heptauthrene,[8] Clar's goblet,[9] acenes,[10-11] anthenes,[12] periacenes,[13-15] rhombene[16], etc.

Although great advancements have been made through wet chemistry and on-surface synthesis, two main challenges impede further progress in this field, firstly, screening prospective magnetic candidates from the vast structural diversity of BPHs. While the enumeration of the BPH graphs is a long-standing topic in mathematics[22] and can be traced back to 1968[23], high-throughput property-oriented exploration based on these structures remains blank. The second obstacle lies in how to realize BPHs with high spins. Lieb's theorem[24], which states the ground state spin quantum number (S) is equal to $1/2\,|N_A - N_B|$, where $N_A$ and $N_B$ represent the numbers of sites from the two interpenetrating sublattices (A and B), respectively, is the most well-known method in this regard. However, it is not capable of determining the distribution of sublattice imbalances, and cannot predict spin-polarized BPHs with open-shell singlet ground states[6, 9].

In this work, we prepared a database of BPHs by enumerating their structures. Through tight-binding (TB) simulation, we explored a new rule, the Triangle Counting Rule (TCR), to unravel the impact of topology on the open-shell characters of BPHs. The TCR is based on and extends the well-established Lieb's theorem and Ovchinnikov's rule. We established a database of all spin multiplet BPHs with less than 10 hexagons through permutation-combination. Through the application of machine learning (ML) technique, we uncovered a strong relationship between the internal vertices of the BPH graphs and their open-shell characters. Based on this topology-magnetism relationship, we simplified Lieb's theorem and established a set of simple rules to predict the magnetic properties of BPHs. Our results are further validated through analysis of several types of open-shell BPHs using both TB and Mean-field Hubbard (MFH) model. Our results not only provide a more straightforward approach to predict the quantum spin number of BPHs, but also reveal the distribution of sublattice imbalance and the spin-polarization of BPHs.

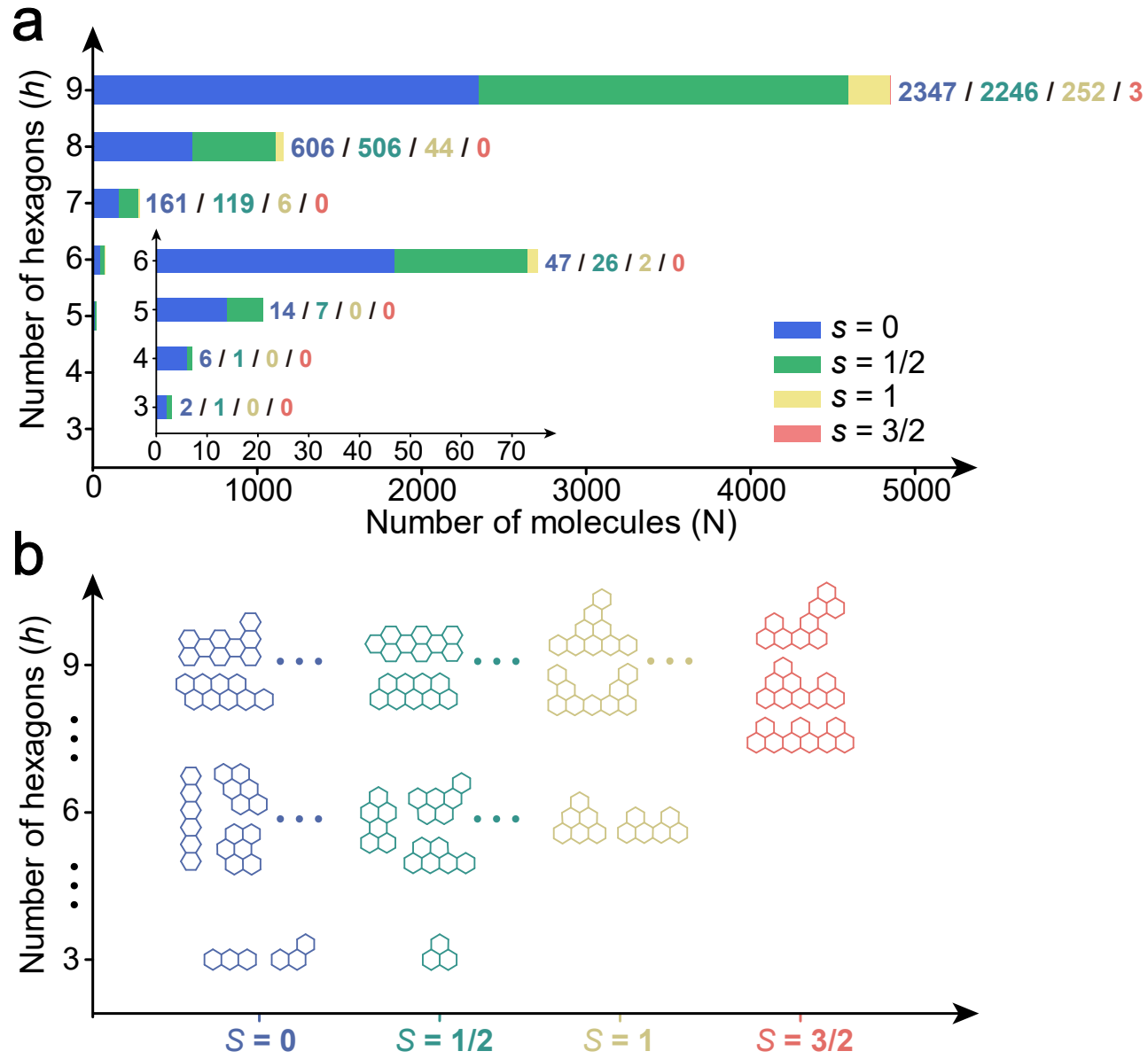

Figure 1. Enumeration of benzenoid polycyclic hydrocarbons (BPHs). (a) Numbers of BPHs fused with $h$ hexagons ($h$ up to 9). Helicenes and coronoids are not included. The colors indicate the numbers of the total spin quantum number S (S = 0, blue; S = 1/2, green; S = 1, yellow; S = 3/2, red). (b) A few examples of the BPH graphs with different S and $h$.

## RESULTS AND DISCUSSION

### Structure database of BPHs

In order to establish a molecular database, we limited the number of hexagonal rings ($h$) in BPHs to less than 10 and excluded helicenes and coronoids, considering only simply connected planar polyhexes. We computationally enumerated all the possible combinations of hexagonal rings through the permutation-combination and screened out duplicated structures with the same structural features (Supporting section 5.1). Our results, which include the numbers (N) of BPHs composed of $h$ hexagons, are consistent with the previous finding (N = 81 when $h$ = 6; N = 331 when $h$ = 7; N=1435 when $h$ = 8, N = 6505 when $h$ = 9),[25] thus validating the reliability of our method (Fig. 1a and Table S1). In the summary plot of Fig. 1a, we removed the structures with the Fjord edges which have two vertices in the closest proximity but are disconnected (Fig. S11), as they are not considered in our analysis due to their highly non-planar configurations. (Complete data can be found in Supporting section 1 and Additional materials)

The magnetic properties of graphene-based nanostructures are intimately related to the appearance of zero-energy states and how they are affected by electron-electron interactions.[24, 26-27] Within the nearest-neighbor TB model, the existence of zero-energy states is correlated with the existence of a sublattice imbalance[26, 28] (non-Kekulean BPHs, an earlier rule that illustrates the presence of zero-energy states with balanced sublattice only emerge at h ⩾ 11)[29] (see Supporting section 5.2 for theoretical details). Therefore, we screened all BPHs with zero energy using the TB model and divided them into different groups with different quantum spin number $S$ according to Lieb's theorem (Fig. 1 and table S1). Our statistics reveal two smallest BPHs with S = 1 when $h$ = 6 and three smallest BPHs with S = 3/2 when $h$ = 9 (see structures in Fig. 1b). Interestingly, with the increase of $h$, the proportion of the BPHs with S ⩾ ½ rises from 31.8% for $h$ = 5 to 50.5% for $h$ = 9. Given that a plethora of open-shell singlet BPHs is not counted (see below), open-shell BPHs account for a major diversity. In Fig. 1b, we present a selection of BPHs classified by their S. For a comprehensive list of open-shell BPH structures with $h$ ⩽ 9, please refer to Supporting section 1.

### Correlation between structural and open-shell characteristics

The integration of big data and machine learning methods has led to remarkable progress in the development of new materials and a deeper understanding of their structure-property relationships.[30-31] To investigate the potential correlation between the structural and open-shell characteristics of BPHs, we employed an interpretable strategy that combines machine learning algorithms with the SHAP approach[32] (see Supporting section 5.6). Our dataset comprises a total of 6392 structures, including 3217 of multiplet characters (S ≥ ½) and 3175 BPHs of singlet characters (S = 0). By taking 80% of them as the training set and 20% as the testing set, we examined the impact of several structural features (all features are explained in Table S2 and Fig. S5) on open-shell characters through four different machine learning classification algorithms (Random Forest, MLP Classifier, Logistic Regression, Support Vector Machines).

The SHAP analysis is utilized to determine the key factors that contribute to the open-shell characters. As shown in the SHAP dot diagram in Fig. 2a, we notice that feature 1, sublattice imbalance, and feature 2, the number of internal vertices of BPH graphs ($N_i$, see examples in Fig. 2b and Fig. S5), have the highest influences on the open-shell properties. By further exploring the correlation between the sublattice imbalance and the number of internal vertices, we discovered that for any BPH, the total number of edge vertex ($N_e$) will always be even (see more details in Supporting section 5.4 and Fig. S12). Therefore, the sublattices of the edge vertex must be balanced ($N_A = N_B = n$). As shown in Fig. 2b, the A-B sublattice imbalance of the entire graph ($N_A - N_B$) can be determined as the A-B sublattice imbalance of the internal vertex ($N_A' - N_B'$):

$$N_A - N_B = (N_A - n) - (N_B - n) = N_A' - N_B'$$

A closer inspection of the topological graph reveals that each internal vertex corresponds to one phenalenyl sub-graph. As a result, we can calculate the total quantum spin number S by counting the number of triangles with different orientations (Fig. 2b, right). This counting rule is significantly simpler and more intuitive than the widely used Lieb's theorem. Additionally, the impact of adding a hexagon to the original structure on its magnetic properties can be deduced by analyzing the changes in its internal sublattice. Fig. 2c demonstrates that adding a hexagon by sharing one edge (from i to ii) does not alter the number of A' and B' sublattices. Adding a hexagon by sharing three or five edges (iii to iv, v to vi), the number of A' and B' sublattices will increase by 1 (three edges) or 2 (five edges), so the total spin number of the molecule will not change. But adding a new hexagon by sharing two or four edges (iii, v) will introduce sublattice imbalance and the S will be changed. Therefore, we only need to judge the position of the newly added hexagon to know the change of the open-shell character of the system.

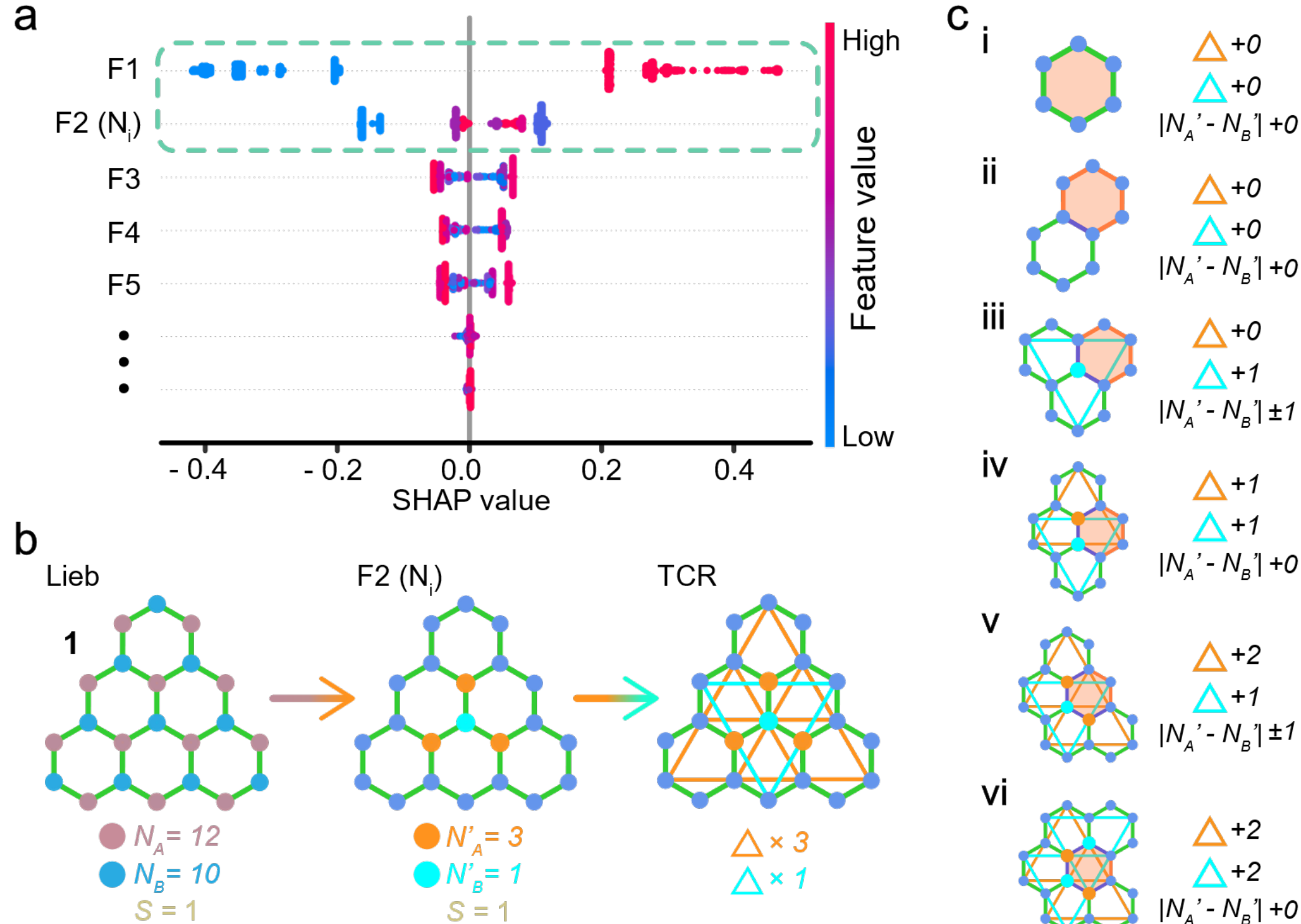

Figure 2. From Lieb's theorem to TCR. (a) The SHAP summary plot shows the impacts of different structural features on the open-shell characters of BPHs. A larger absolute SHAP value means a higher correlation between the feature and the magnetism predicted by the ML model. (b) Left: The sublattice imbalance and the resulting total spin from Lieb's theorem for [3]triangulene. brown and blue circles represent A and B sublattice atoms, respectively. Middle: calculating the total spin by considering only the internal vertices. Yellow and cyan circles represent A' and B' sublattice atoms, respectively. Right: the triangular counting rule. Each interior vertex corresponds to a phenalenyl triangle with a certain orientation. (c) Variation of the number of phenalenyl triangles and sublattice imbalances by adding hexagons on the BPH in six different possible ways. The hexagon filled in red represents newly added hexagons. The internal vertex A' and B' and the corresponding triangles are colored orange and cyan, respectively.

### The triangle counting rule (TCR)

If we apply this rule to larger triangles, we find that when the order of the triangle increases by 1, the spin quantum number increases by 1/2. As displayed in Fig. 3c, three phenalenyl triangles with the same orientation highlighted in orange and one phenalenyl triangle with the opposite orientation in cyan are found in [3]-triangulene[3] structure **1** leading to S = 1, while six orange and three cyan phenalenyl triangles are found in [4]-triangulene[2] structure **2** leading to S = 3/2. The correspondence between the internal vertex and triangle is useful since the triangle includes information of orientations. Therefore, the total spin number $S$ can be determined by counting the number of triangles with different orientations. For instance, two orange phenalenyl triangle can be identified from heptauthrene **3**[8] leading to S = 1, whereas **4** with two orange phenalenyl triangle and one cyan phenalenyl triangle has S = 1/2 (Fig. 3b). In some cases, if we use higher-order triangles instead of lots of lower-order triangles, the calculation will be much simpler, such as the bipartite nanographene (BNG) **5**[17] and the extended triangulene **6**[19]. Finally, this rule also could screen some BPHs with singlet ground state, for example, the well-known super-heptazethrene **7**[7] and Clar's goblet 8[9]. Most of the molecules discussed above have been studied either on-surface or in solution, which further supports our results.

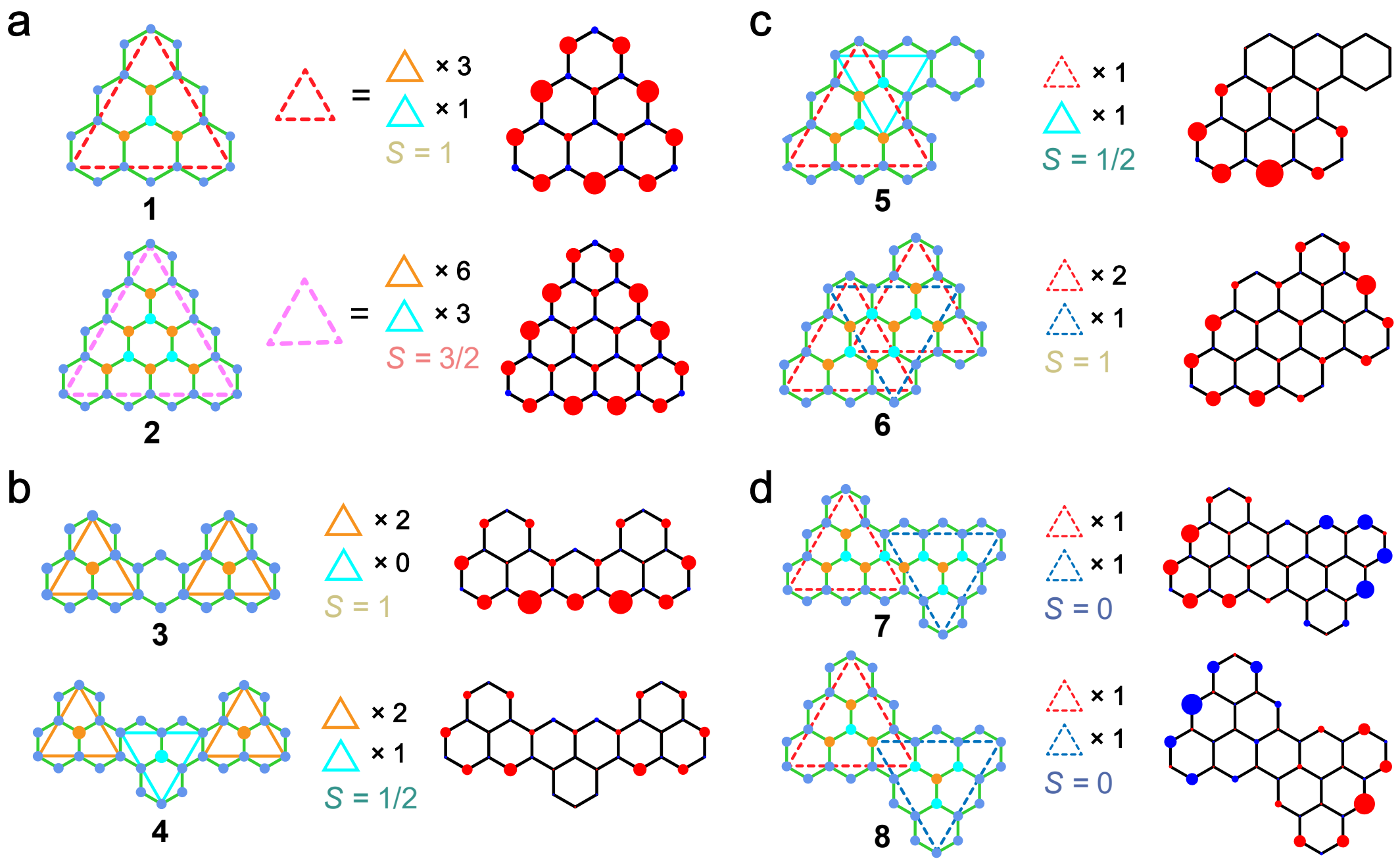

Figure 3. Illustration of the triangle counting rule (TCR). (a) Counting the phenalenyl triangles in high-order molecular triangles. (b) Two BPHs with multiple phenalenyl triangles, the magnetic ground state of the BPHs can be identified by the number of two kinds of triangles with different orientations. (c) Two BPHs in which the phenalenyl triangles are overlapped with each other. (d) Two BPHs with open-shell singlet ground state have the same number of different triangles. In (a-d), the MFH calculated spin density distributions of the corresponding BPH structures are shown on the right. Red/blue isosurfaces denote opposite signs of the spin distributions.

Besides the magnetic properties originating from the zero-energy state, some BPHs exhibit open-shell properties due to the spin-polarization of low-energy states or competition between the aromatic stabilization and although the total spin S of the molecule equals 0, penalty of forming radicals.[33] However, the prediction of these open-shell singlets is more challenging than the multiplet species because it is quite difficult to estimate the degree of polarization of a BPH without detailed calculations. The TCR method can be applied to predict this kind of magnetism. We can judge the distribution of spin units by TCR (each triangle can be regarded as a region with local spin), and the key factor for judging the open-shell/closed-shell singlet is the length of the 'bridge' between spin units in different directions. For instance, zethrenes can be viewed as two oppositely oriented triangles connected by varying the number of bridging hexagons. We find that zethrenes develop increasing open-shell characters with the elongation of the central bridge (Fig. 4a) as demonstrated by the absolute magnetization from the MFH model. We have computed different zethrene derivatives with up to 3 bridging hexagons (all structures are shown in Fig. S4) and we find that the threshold number of the bridging hexagons to give rise to open-shell singlet is 2. To support our findings, we also calculated their diradical character $y_0$, the singlet polarization degree, from 0 to 1, corresponding to absolute closed-shell and diradical states, respectively[34] (Details in Supporting section 5.5).

There is another kind of BPH with open-shell singlet ground state but their magnetic properties derive from the topological frustration.[9, 35] Fig. 4b-i shows the structure of one of such BPHs, Clar's goblet **8**, along with its computed spin density map. the system's nullity amounts to $\eta$=2. This is corroborated by TB calculations, which predict the presence of two zero-energy states (Fig. 4b-i). The zero-energy states become spin-polarized as Coulomb interactions are included with the Hubbard model (Fig. 4b-i). By drawing the phenalenyl triangles, we find that such topologically frustrated BPHs could be considered as two BPHs of opposite spin moments connected by a phenyl bridge with no spin density distributions (Fig. 4b, right). The diradical characters $y_0$ of this kind of BPH are always 1, further demonstrating their pure diradical property. We find topological frustrated BPHs require at least eleven hexagons (six for the two phenalenyl triangles and five for the perylene bridge). This is consistent with the fact that in our BPH database with h less than 10 we could not find any singlet molecules with zero-energy states. Based on this design rule, we conceived seven molecules of topological frustration comprising 11 hexagons (Fig. 4b and Fig. S10).

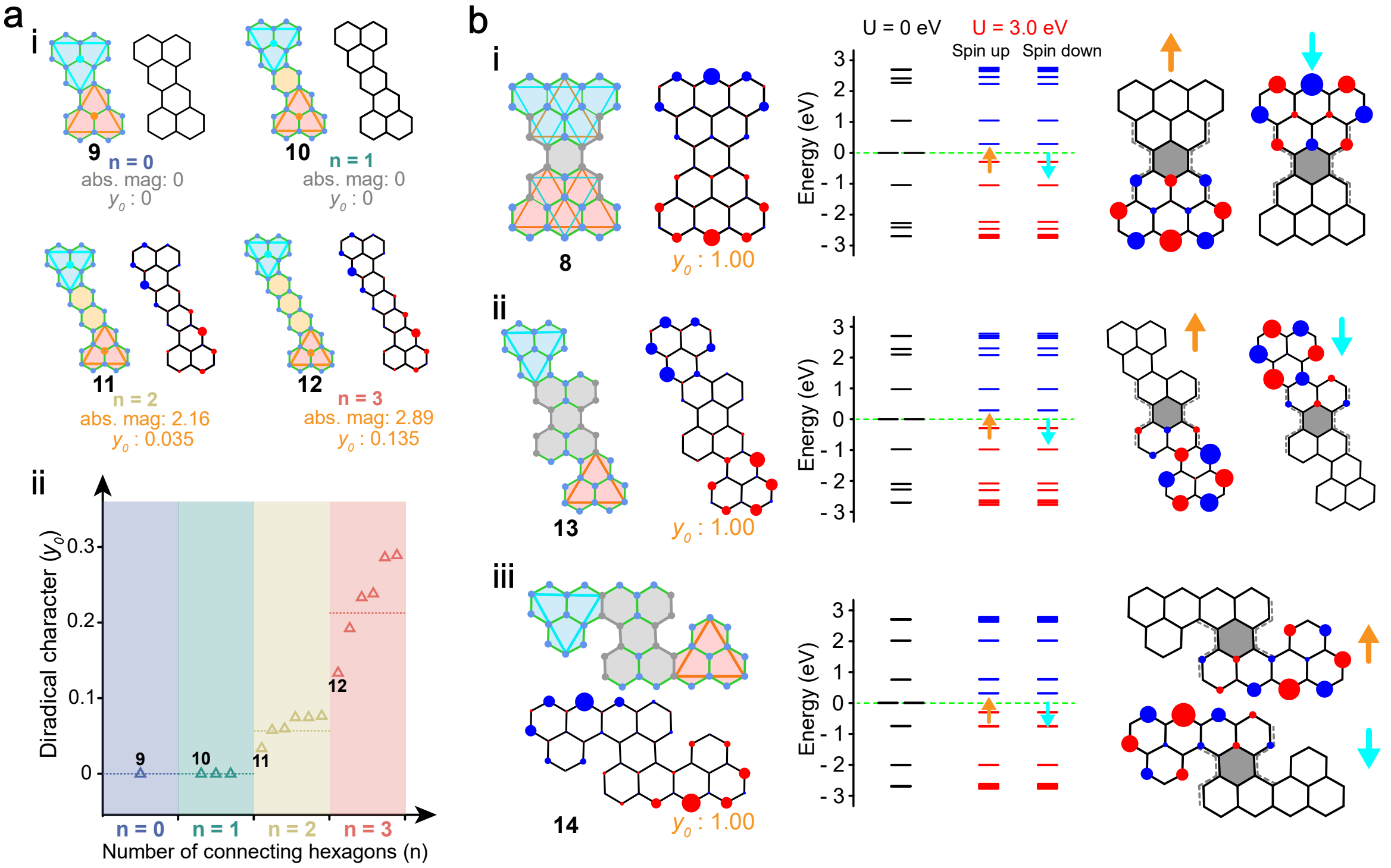

Figure 4. Applying the TCR to singlet BPHs. (a) Zethrenes. (a-i) A series of zethrene with an increasing number of connecting hexagons (n) and corresponding spin density maps. (a-ii) Summary of the diradical character $y_0$ with the increasing number of connecting hexagons. (b) BPHs with topological frustrations. (b-i) Left: the structure of Clar's goblet **8** and corresponding spin density distribution. Middle: the TB and MFH computed energy spectra of **8**. Right: MFH-calculated wave functions of two SOMOs. (b-ii), (b-iii) Two additional BPHs with topological frustration. Red/blue isosurfaces denote opposite signs of the spin direction or wave function.

## CONCLUSION

To conclude, we have built up a structure database of BPHs with up to 9 hexagons by graph enumeration and screened the open-shell BPHs by computing the TB eigenvalues of the BPH graphs, which may serve to guide the experimental realizations of BPHs with desired magnetic properties. Through a combination of data-driven analysis and an interpretable SHAP approach, we have unveiled the correlation between the internal vertices of the BPH graph and their open-shell characters. We have also introduced the triangle counting rule (TCR) as an alternative approach to predict the magnetic ground states of BPHs. This method simplifies and extends the well-known Lieb's theorem by: (1) considering only the internal vertex number; (2) treating each triangle as a spin unit for spin density distribution; (3) providing predictions into the open-shell properties of S = 0 BPHs with the phenalenyl subgraphs. With the recent advances in emerging synthetic routes,[36-37] we foresee the experimental realizations of these predicted structures and further applications of them in further quantum technologies.

## ASSOCIATED CONTENT

### Supporting Information

The Supporting Information is available free of charge on the ACS Publications website at DOI:

## AUTHOR INFORMATION

### Corresponding Author

*qiangsun@shu.edu.cn

### Author Contributions

Y.Y. , B.Q. and Q.S. wrote the paper. F.Z. did the ML work and Y.Y performed quantum chemistry calculations. All authors discussed the results and commented on the manuscript at all stages.

### Notes

The authors declare no competing financial interests.

## ACKNOWLEDGMENT

This work was supported by National Natural Science Foundation of China (No. 22072086). We thank Fuyi Wang at Nanjing University for his helps in coding.

## REFERENCES

1. Tombros, N.; Jozsa, C.; Popinciuc, M.; Jonkman, H. T.; van Wees, B. J., Electronic spin transport and spin precession in single graphene layers at room temperature. *Nature* **2007,** *448* (7153), 571-574.
2. Mishra, S.; Beyer, D.; Eimre, K.; Liu, J.; Berger, R.; Gröning, O.; Pignedoli, C. A.; Müllen, K.; Fasel, R.; Feng, X.; Ruffieux, P., Synthesis and Characterization of π-Extended Triangulene. *J. Am. Chem. Soc.* **2019,** *141* (27), 10621-10625.
3. Pavlicek, N.; Mistry, A.; Majzik, Z.; Moll, N.; Meyer, G.; Fox, D. J.; Gross, L., Synthesis and characterization of triangulene. *Nat Nanotechnol* **2017,** *12* (4), 308-311.
4. Su, J.; Fan, W.; Mutombo, P.; Peng, X.; Song, S.; Ondráček, M.; Golub, P.; Brabec, J.; Veis, L.; Telychko, M.; Jelínek, P.; Wu, J.; Lu, J., On-Surface Synthesis and Characterization of [7]Triangulene Quantum Ring. *Nano Lett.* **2021,** *21* (1), 861-867.
5. Su, J.; Telychko, M.; Hu, P.; Macam, G.; Mutombo, P.; Zhang, H.; Bao, Y.; Cheng, F.; Huang, Z.-Q.; Qiu, Z.; Tan, S. J. R.; Lin, H.; Jelínek, P.; Chuang, F.-C.; Wu, J.; Lu, J., Atomically precise bottom-up synthesis of π-extended [5]triangulene. **2019,** *5* (7), eaav7717.

6. Turco, E.; Mishra, S.; Melidonie, J.; Eimre, K.; Obermann, S.; Pignedoli, C. A.; Fasel, R.; Feng, X.; Ruffieux, P., On-Surface Synthesis and Characterization of Super-nonazethrene. *The Journal of Physical Chemistry Letters* **2021,** *12* (34), 8314-8319.

7. Mishra, S.; Melidonie, J.; Eimre, K.; Obermann, S.; Gröning, O.; Pignedoli, C. A.; Ruffieux, P.; Feng, X.; Fasel, R., On-surface synthesis of super-heptazethrene. *Chem. Commun.* **2020,** *56* (54), 7467-7470.

8. Su, X.; Li, C.; Du, Q.; Tao, K.; Wang, S.; Yu, P., Atomically Precise Synthesis and Characterization of Heptauthrene with Triplet Ground State. *Nano Lett.* **2020,** *20* (9), 6859-6864.

9. Mishra, S.; Beyer, D.; Eimre, K.; Kezilebieke, S.; Berger, R.; Gröning, O.; Pignedoli, C. A.; Müllen, K.; Liljeroth, P.; Ruffieux, P.; Feng, X.; Fasel, R., Topological frustration induces unconventional magnetism in a nanographene. *Nature Nanotechnology* **2020,** *15* (1), 22-28.

10. Urgel, J. I.; Mishra, S.; Hayashi, H.; Wilhelm, J.; Pignedoli, C. A.; Di Giovannantonio, M.; Widmer, R.; Yamashita, M.; Hieda, N.; Ruffieux, P.; Yamada, H.; Fasel, R., On-surface light-induced generation of higher acenes and elucidation of their open-shell character. *Nature Communications* **2019,** *10* (1), 861.

11. Zuzak, R.; Dorel, R.; Kolmer, M.; Szymonski, M.; Godlewski, S.; Echavarren, A. M., Higher Acenes by On-Surface Dehydrogenation: From Heptacene to Undecacene. *Angew. Chem. Int. Ed.* **2018,** *57* (33), 10500-10505.

12. Wang, S.; Talirz, L.; Pignedoli, C. A.; Feng, X.; Mullen, K.; Fasel, R.; Ruffieux, P., Giant edge state splitting at atomically precise graphene zigzag edges. *Nat Commun* **2016,** *7*, 11507.

13. Rogers, C.; Chen, C.; Pedramrazi, Z.; Omrani, A. A.; Tsai, H.-Z.; Jung, H. S.; Lin, S.; Crommie, M. F.; Fischer, F. R., Closing the Nanographene Gap: Surface-Assisted Synthesis of Peripentacene from 6,6′ -Bipentacene Precursors. *Angew. Chem. Int. Ed.* **2015,** *54* (50), 15143-15146.

14. Sánchez-Grande, A.; Urgel, J. I.; Veis, L.; Edalatmanesh, S.; Santos, J.; Lauwaet, K.; Mutombo, P.; Gallego, J. M.; Brabec, J.; Beran, P.; Nachtigallová, D.; Miranda, R.; Martín, N.; Jelínek, P.; Écija, D., Unravelling the Open-Shell Character of Peripentacene on Au(111). *The Journal of Physical Chemistry Letters* **2021,** *12* (1), 330-336.

15. Mishra, S.; Lohr, T. G.; Pignedoli, C. A.; Liu, J.; Berger, R.; Urgel, J. I.; Müllen, K.; Feng, X.; Ruffieux, P.; Fasel, R., Tailoring Bond Topologies in Open-Shell Graphene Nanostructures. *ACS Nano* **2018,** *12* (12), 11917-11927.

16. Mishra, S.; Yao, X.; Chen, Q.; Eimre, K.; Gröning, O.; Ortiz, R.; Di Giovannantonio, M.; Sancho-García, J. C.; Fernández-Rossier, J.; Pignedoli, C. A.; Müllen, K.; Ruffieux, P.; Narita, A.; Fasel, R., Large magnetic exchange coupling in rhombus-shaped nanographenes with zigzag periphery. *Nature Chemistry* **2021,** *13* (6), 581-586.

17. Zheng, Y.; Li, C.; Xu, C.; Beyer, D.; Yue, X.; Zhao, Y.; Wang, G.; Guan, D.; Li, Y.; Zheng, H.; Liu, C.; Liu, J.; Wang, X.; Luo, W.; Feng, X.; Wang, S.; Jia, J., Designer spin order in diradical nanographenes. *Nature Communications* **2020,** *11* (1), 6076.

18. Zheng, Y.; Li, C.; Zhao, Y.; Beyer, D.; Wang, G.; Xu, C.; Yue, X.; Chen, Y.; Guan, D.-D.; Li, Y.-Y.; Zheng, H.; Liu, C.; Luo, W.; Feng, X.; Wang, S.; Jia, J., Engineering of Magnetic Coupling in Nanographene. *Phys. Rev. Lett.* **2020,** *124* (14), 147206.

19. Li, J.; Sanz, S.; Castro-Esteban, J.; Vilas-Varela, M.; Friedrich, N.; Frederiksen, T.; Pena, D.; Pascual, J. I., Uncovering the Triplet Ground State of Triangular Graphene Nanoflakes Engineered with Atomic Precision on a Metal Surface. *Phys. Rev. Lett.* **2020,** *124* (17), 177201.

20. Schlenk, W.; Brauns, M., Über einige Bis-triarylmethyle. [Über Triarylmethyle. XV.]. *Berichte der deutschen chemischen Gesellschaft* **1915,** *48* (1), 716-728.

21. Rajca, A., Organic Diradicals and Polyradicals: From Spin Coupling to Magnetism? *Chem. Rev.* **1994,** *94* (4), 871-893.

22. Gutman, I.; Cyvin, S. J. In *Introduction to the theory of benzenoid hydrocarbons*, 1989.

23. Balaban, A. T.; Harary, F., Chemical graphs—V: Enumeration and proposed nomenclature of benzenoid cata-condensed polycyclic aromatic hydrocarbons. *Tetrahedron* **1968,** *24* (6), 2505-2516.

24. Lieb, E. H., Two theorems on the Hubbard model. *Phys. Rev. Lett.* **1989,** *62* (10), 1201-1204.

25. Vöge, M.; Guttmann, A. J.; Jensen, I., On the Number of Benzenoid Hydrocarbons. *Journal of Chemical Information and Computer Sciences* **2002,** *42* (3), 456-466.

26. Inui, M.; Trugman, S. A.; Abrahams, E., Unusual properties of midband states in systems with off-diagonal disorder. *Physical Review B* **1994,** *49* (5), 3190-3196.

27. Fajtlowicz, S.; John, P.; Sachs, H., On Maximum Matchings and Eigenvalues of Benzenoid Graphs. *Croatica Chemica Acta (CCA@chem.pmf.hr); Vol.78 No.2* **2005,** *78*.

28. Fernández-Rossier, J.; Palacios, J. J., Magnetism in Graphene Nanoislands. *Phys. Rev. Lett.* **2007,** *99* (17), 177204.

29. Cyvin, S. J.; Brunvoll, J.; Cyvin, B. N., The hunt for concealed non-Kekuléan polyhexes. *J. Math. Chem.* **1990,** *4* (1), 47-54.

30. Butler, K. T.; Davies, D. W.; Cartwright, H.; Isayev, O.; Walsh, A., Machine learning for molecular and materials science. *Nature* **2018,** *559* (7715), 547-555.

31. Behler, J., Perspective: Machine learning potentials for atomistic simulations. *J. Chem. Phys.* **2016,** *145* (17), 170901.

32. Budescu, D. V., Dominance analysis: A new approach to the problem of relative importance of predictors in multiple regression. *Psychological Bulletin* **1993,** *114* (3), 542-551.

33. Oteyza, D.; Frederiksen, T., *Carbon-based nanostructures as a versatile platform for tunable $\pi$-magnetism*. 2022.

34. Nakano, M.; Kishi, R.; Nakagawa, N.; Ohta, S.; Takahashi, H.; Furukawa, S.-i.; Kamada, K.; Ohta, K.; Champagne, B.; Botek, E.; Yamada, S.; Yamaguchi, K., Second Hyperpolarizabilities (γ) of Bisimidazole and Bistriazole Benzenes: Diradical Character, Charged State, and Spin State Dependences. *The Journal of Physical Chemistry A* **2006,** *110* (12), 4238-4243.

35. Wang, W. L.; Yazyev, O. V.; Meng, S.; Kaxiras, E., Topological Frustration in Graphene Nanoflakes: Magnetic Order and Spin Logic Devices. *Phys. Rev. Lett.* **2009,** *102* (15), 157201.

36. Grill, L.; Hecht, S., Covalent on-surface polymerization. *Nature Chemistry* **2020,** *12* (2), 115-130.

37. Kinikar, A.; Di Giovannantonio, M.; Urgel, J. I.; Eimre, K.; Qiu, Z.; Gu, Y.; Jin, E.; Narita, A.; Wang, X.-Y.; Müllen, K.; Ruffieux, P.; Pignedoli, C. A.; Fasel, R., On-surface polyarylene synthesis by cycloaromatization of isopropyl substituents. *Nature Synthesis* **2022,** *1* (4), 289-296.